\documentclass[%
 aip,
 amsmath,amssymb,
 reprint,%
]{revtex4-1}

\usepackage{graphicx}% Include figure files
\usepackage{dcolumn}% Align table columns on decimal point
\usepackage{bm}% bold math
\usepackage{booktabs}
\usepackage[utf8]{inputenc}
\usepackage[T1]{fontenc}
\usepackage{mathptmx}
\usepackage{etoolbox}
\usepackage{color}
\makeatletter
\def\@email#1#2{%
 \endgroup
 \patchcmd{\titleblock@produce}
  {\frontmatter@RRAPformat}
  {\frontmatter@RRAPformat{\produce@RRAP{*#1\href{mailto:#2}{#2}}}\frontmatter@RRAPformat}
  {}{}
}%
\makeatother
\begin{document}

\preprint{AIP/123-QED}

\title{Beam-Tracing-Based Quantitative Reconstruction of Density Fluctuations in QUEST Using Doppler Backscattering}

\author{T. Kinoshita}
\email{t.kinoshita@triam.kyushu-u.ac.jp}
\affiliation{Research Institute for Applied Mechanics, Kyushu University, Kasuga, Fukuoka 816-8580, Japan}

\author{T. Tokuzawa}
\affiliation{National Institute for Fusion Science, Toki, Gifu 509-5292, Japan}

\author{V. H. Hall-Chen}
\author{Y. T. Tan}
\affiliation{Future Energy Acceleration \& Translation (FEAT), Agency for Science, Technology and Research (A*STAR), Singapore 138632, Singapore}

\author{T. Ido}
\affiliation{Research Institute for Applied Mechanics, Kyushu University, Kasuga, Fukuoka 816-8580, Japan}
\affiliation{National Institute for Fusion Science, Toki, Gifu 509-5292, Japan}

\author{H. Idei}
\author{R. Ikezoe}
\author{K. Hanada}
\author{M. Hasegawa}
\author{T. Onchi}
\author{Y. Peng}
 \affiliation{Research Institute for Applied Mechanics, Kyushu University, Kasuga, Fukuoka 816-8580, Japan}

\author{QUEST Experimental Group}

\date{\today}% It is always \today, today,
\begin{abstract}
A three-channel X-/Ku-band Doppler backscattering (DBS) system has been developed and installed on QUEST for turbulence and electric-field measurements. 
In spherical tokamaks, the large magnetic-field pitch angle increases the geometric mismatch between the probing beam wave vector and the local magnetic-field vector, reducing the effective perpendicular projection and resulting in a systematic underestimation of the measured scattering intensity. 
In addition, in QUEST, where low plasma density requires a low-frequency probe beam, beam propagation effects become increasingly significant, further complicating the interpretation of the measured DBS power in terms of local density fluctuation amplitude.
To address these issues, a quantitative correction methodology based on the synthetic DBS code SCOTTY was established. 
All relevant diagnostic response effects were evaluated using SCOTTY along ray trajectories, yielding a correction factor for reconstructing the local turbulence amplitude from the measured scattering signal.
The correction factor exhibits strong spatial and frequency dependence, varying by up to an order of magnitude between the plasma core and edge regions, highlighting the necessity of frequency-dependent corrections.
By applying the derived correction factor to experimental measurements, quantitative density fluctuation amplitudes were reconstructed from the detected scattering signals.
Evaluation of the fluctuation amplitude indicates enhanced turbulence activity in the plasma edge region, where a finite negative radial electric field is inferred.
This work demonstrates the first quantitative turbulence evaluation using low-frequency X-/Ku-band DBS measurements in QUEST and establishes a framework for quantitative DBS analysis in spherical tokamaks.
\end{abstract}

\maketitle

\section{\label{sec:intro}Introduction}
Doppler backscattering (DBS), also referred to as Doppler reflectometry, is a powerful microwave diagnostic technique used in magnetic confinement fusion devices to simultaneously measure the local density fluctuation amplitude $\tilde{n}_e$, perpendicular wavenumber $k_\perp$, and plasma rotation velocity $v_\perp$ from the Doppler frequency shift, $f_D=v_\perp k_\perp /2\pi$. 

QUEST (Q-shu University Experiment with Steady-State Spherical Tokamak) is a spherical tokamak (ST) dedicated to the development of non-inductive plasma start-up and long-pulse plasma operation \cite{hanada2025experimental}. 
The device has a major radius of $R=0.64$ m, a minor radius of $a=0.42$ m, and operates at a relatively low toroidal magnetic field of up to $0.25$ T.
One of the primary motivations for developing a DBS system on QUEST is to investigate aspect-ratio effects on plasma confinement.
Equilibrium calculations indicate that QUEST can, in principle, access a wide range of plasma aspect ratios ($A \sim 1.6$--2.5), enabling configurations from ST-like to conventional tokamak-like geometries.
Since significant differences in confinement scaling have been reported between conventional and spherical tokamaks \cite{doyle2007chapter,kaye2021thermal}, DBS measurements on QUEST provide a unique opportunity to study the role of turbulence in aspect-ratio-dependence.
The typical electron density in QUEST plasmas sustained by electron cyclotron current drive (ECCD) start-up and electron cyclotron resonance heating (ECRH) is on the order of $10^{18}~\mathrm{m^{-3}}$, requiring DBS measurements in the X- and Ku-band frequency ranges.
While low-frequency reflectometry systems have been developed in linear plasma devices such as GAMMA 10 \cite{kohagura2022ku}, the application of X-/Ku-band DBS to toroidal magnetic confinement devices has rarely been reported.
In low-density spherical tokamaks, the low electron density necessitates the use of long probing wavelengths, while the spherical tokamak geometry introduces strong magnetic geometry effects. 
Together, these factors can significantly modify the scattering localization and efficiency.
As a result, the measured DBS power cannot be directly interpreted as the local density fluctuation amplitude.
%In conventional analyses, the scattered signal intensity is often assumed to be proportional to the local turbulence intensity, i.e., $\delta n_e^2$, under the assumption of weak variations in the diagnostic response function.

This paper presents both the development of a three-channel X-/Ku-band DBS system and the establishment of a beam-tracing-based quantitative analysis methodology using the SCOTTY code\cite{hall2022beam}.
The diagnostic response, including beam propagation and scattering geometry effects, is evaluated to derive correction factors for the measured DBS power.
The proposed approach enables a more quantitative evaluation of density fluctuation amplitudes in spherical tokamak plasmas, particularly in the low-density regime.

\section{\label{sec:system}Development of QUEST X-/Ku-band DBS System}
DBS is a microwave diagnostic technique used to measure plasma density fluctuations and their perpendicular propagation velocity by detecting the Doppler frequency shift of waves backscattered near the plasma cutoff layer.
The fluctuation wavenumber selected by the diagnostic is determined by the Bragg scattering condition, which relates the fluctuation wavenumber to the local microwave wavenumber at the scattering location.
Although the Bragg condition may be satisfied at multiple locations along the beam trajectory, the detected signal is strongly localized near the cutoff layer, where the scattering efficiency is maximized due to the enhanced wave electric field.
Therefore, varying the probing frequency changes the cutoff position and hence the measurement location, while varying the beam injection angle changes the Bragg condition and thereby the selected fluctuation wavenumber.

\begin{figure}[t]
\centering
\includegraphics[width=\columnwidth]{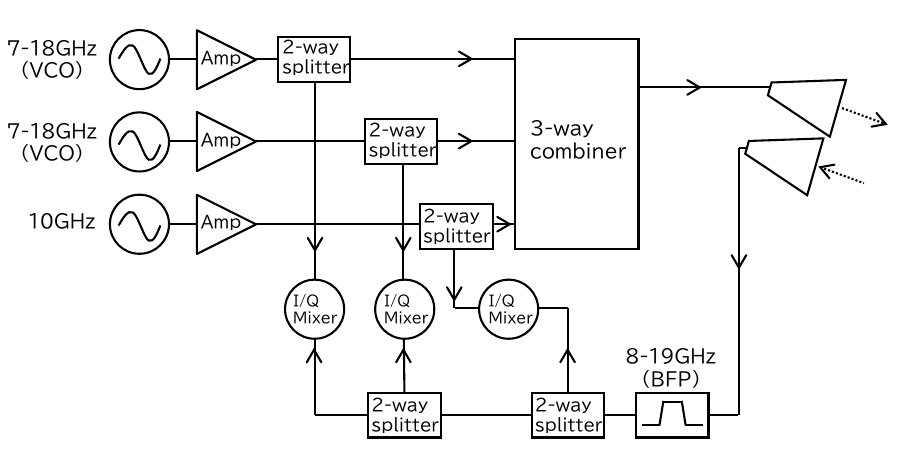}
\caption{Schematic diagram of the X-/Ku-band DBS system installed on QUEST. The isolators used in the transmission lines are omitted for clarity.}
\label{fig:DBSsys}
\end{figure}

Figure \ref{fig:DBSsys} shows the schematic diagram of the X-/Ku-band DBS system developed for QUEST.
The target plasmas are 28~GHz ECRH plasmas, in which the electron density is typically on the order of $10^{18}~\mathrm{m^{-3}}$.
To access the corresponding O-mode cutoff layers, probing frequencies in the X- and Ku-band ranges were selected.
The DBS system consists of three independent channels employing one fixed-frequency 10~GHz source and two voltage-controlled oscillators (VCOs) covering 7--18~GHz, enabling simultaneous measurements at different radial locations.
Before transmission, each probing microwave is divided into two paths.
One path is launched into the plasma through the optical system, while the other is supplied to the LO port of the corresponding I$/$Q mixer as a reference signal.
In the QUEST configuration, the plasma is located 0.7--1.0~m from the vacuum vessel window.
Owing to this long propagation distance and the relatively low probing frequencies, the microwave beam undergoes significant divergence.
To compensate for this effect, a focusing optical system based on Gaussian beam optics was designed to focus the beam near the cutoff layer, thereby maximizing the received backscattered power.
The optical system consists of two mirrors with focal lengths of 350~mm and 400~mm.
Laboratory beam characterization at 14~GHz confirmed a beam waist diameter of approximately 120~mm at the designed focal position.
In addition, the poloidal launch angle can be varied, enabling measurements at different $k_\perp$ values for evaluation of the fluctuation wavenumber spectrum.
The backscattered signal received by the antenna contains all three probing-frequency components and is first passed through a band-pass filter (BPF) to suppress stray radiation from the 28~GHz ECRH system.
The filtered signal is then divided into three identical paths, each of which is fed to the RF port of an I$/$Q mixer.
Since each I$/$Q mixer is driven by the reference signal supplied to its LO port, only the RF component whose frequency matches that of the LO is directly downconverted to baseband, whereas the remaining components are converted to non-zero intermediate frequencies.

\section{\label{sec:method}Theoretical Model and Correction Methodology}
To evaluate the density fluctuation amplitude from the detected backscattered signals, we adopt the analytical beam model derived by Hall-Chen \textit{et al.} \cite{hall2022beam}. 
In this framework, the backscattered power spectral density $p_r(\omega)$ normalized by the launched power $P_{\rm ant}$ is expressed as a line integral along the central ray path $l$. 
The normalized backscattered power spectral density can be written as:
\begin{widetext}
\begin{equation}
\frac{p_r(\omega)}{P_{\rm ant}}
\propto
\frac{1}{\Omega^2 \bar{W}_y}
\int
\varepsilon
\frac{g_{\rm ant}^2}{g^2}
\frac{
\bar{W}_y \det [\mathrm{Im}(\Psi_w)]
}
{
\sqrt{2} |\det(M_w)| [-\mathrm{Im}(M_{yy}^{-1})]^{1/2}
}
\exp\!\left(
-2\frac{\theta_m^2}{(\Delta\theta_m)^2}
\right)
\left|
\delta \tilde{n}_e
\!\left(
k_\perp(l),
\omega
\right)
\right|^2
\,dl
\label{eq:hall_chen_spectral}
\end{equation}

\begin{equation}
\langle \delta n_e^2 (k_\perp)\rangle 
\propto
\frac{P_r}{P_{\rm ant}}
\frac{\Omega^2 \bar{W}_y}
{\displaystyle
\int_{L_1}^{L_2}
\varepsilon
\frac{g_{\rm ant}^2}{g^2}
\frac{
\bar{W}_y \det[\mathrm{Im}(\Psi_w)]
}
{
\sqrt{2} |\det(M_w)| \left[-\mathrm{Im}(M_{yy}^{-1})\right]^{1/2}
}
\exp\!\left(
-2\frac{\theta_m^2}{(\Delta\theta_m)^2}
\right)
\,dl
}
\label{eq:fluctuation_correction}
\end{equation}
\end{widetext}
In Eq.~(\ref{eq:hall_chen_spectral}), $\Omega$ is the probing frequency, and $\bar{W}_y$ is the reference beam waist radius in vacuum.
The polarization factor $\varepsilon$ describes the polarization dependence of the diagnostic response, while the ray factor $g_{\rm ant}^2/g^2$ accounts for the optical amplification associated with the Airy swelling effect near the cutoff layer. 
The beam factor, represented by the terms containing $\bar{W}_y$, $\Psi_w$, and $M_w$, characterizes the effective scattering volume and beam distortion during propagation.
The mismatch term $\exp[-2\theta_m^2/(\Delta\theta_m)^2]$ describes the reduction in scattering efficiency due to the magnetic mismatch angle $\theta_m$ between the probing wavevector and the local magnetic field. 
Finally, $|\delta \tilde{n}_e(k_\perp(l),\omega)|^2$ denotes the density fluctuation spectrum evaluated at the Bragg-matched perpendicular wavenumber $k_\perp(l)$.

The transition from the spectral model in Eq.~(\ref{eq:hall_chen_spectral}) to the inversion formula in Eq.~(\ref{eq:fluctuation_correction}) requires a physical assumption regarding the spatial distribution of turbulence. Since the DBS signal is inherently integrated along the beam trajectory, the local fluctuation intensity cannot be uniquely reconstructed from the received power alone. We define the total received power as $P_r = \int p_r(\omega)\, d\omega$, which corresponds to the area of the Gaussian fit to the Doppler-shifted component after removing the zero-frequency reflection.
We relate this to the density fluctuation intensity at a representative perpendicular wavenumber determined by the Bragg condition, defined as $\langle \delta n_e^2 (k_\perp)\rangle \approx \int \left| \delta \tilde n_e(k_\perp \simeq k_0, \omega) \right|^2 d\omega$ .
Within the effective scattering region, defined as the interval between $L_1$ and $L_2$ where the diagnostic sensitivity function remains above $1/e^2$ of its maximum, we assume that $\langle \delta n_e^2 (k_\perp)\rangle$ is approximately uniform. This assumption is physically motivated by the inherent trade-off between diagnostic localization and turbulence variation in DBS measurements. Regions with weaker density gradients exhibit broader effective scattering regions but generally involve weaker spatial variations in turbulence intensity, whereas steeper density gradients improve localization while allowing for stronger turbulence gradients. Given this trade-off, it is reasonable to approximate $\langle \delta n_e^2 (k_\perp)\rangle$ as constant within the effective scattering volume.
Under this assumption, $\langle \delta n_e^2 (k_\perp)\rangle$ can be taken outside the line integral, allowing the forward model in Eq.~(\ref{eq:hall_chen_spectral}) to be rearranged into the direct inversion form shown in Eq.~(\ref{eq:fluctuation_correction}).
To recover the underlying turbulence level, the measured normalized power $P_r/P_{\rm ant}$ is multiplied by the inverse of the comprehensive correction factor $CF$, where $CF$ is the product of the polarization, ray, beam, and mismatch factors evaluated along the beam path. These diagnostic response factors are evaluated along the beam trajectory using the synthetic DBS code SCOTTY \cite{hall2022beam}, and their profiles under the present experimental conditions are discussed in Sec.~\ref{sec:calibration}.

Unlike the forward-modeling approach of Pratt \textit{et al.} \cite{pratt2024density}, which relies on matching experimental data with synthetic turbulence from gyrokinetic simulations, the present method provides a direct reconstruction based solely on the full diagnostic filter function without requiring an \textit{a priori} turbulence transport model.

\section{Experimental Results and Signal Correction}
\subsection{Experimental Observation}
The target plasmas in this study were hydrogen plasmas initiated by 28~GHz ECCD and sustained by ECRH. To obtain a comprehensive multi-frequency dataset, three highly reproducible discharges (\#55483, \#55484, and \#55485) were selected for analysis. 
During these discharges, the three-channel DBS system was operated simultaneously, with one channel fixed at a probing frequency of 10~GHz, while the probing frequencies of the other two channels were varied between discharges. 
This measurement scheme provided a total of five probing frequencies (8, 9, 10, 13, and 14~GHz). 
The beam launch angles were fixed at $3^\circ$ in the poloidal direction and $0^\circ$ in the toroidal direction throughout the experiments.

\begin{figure}[t]
\centering
\includegraphics[width=\columnwidth]{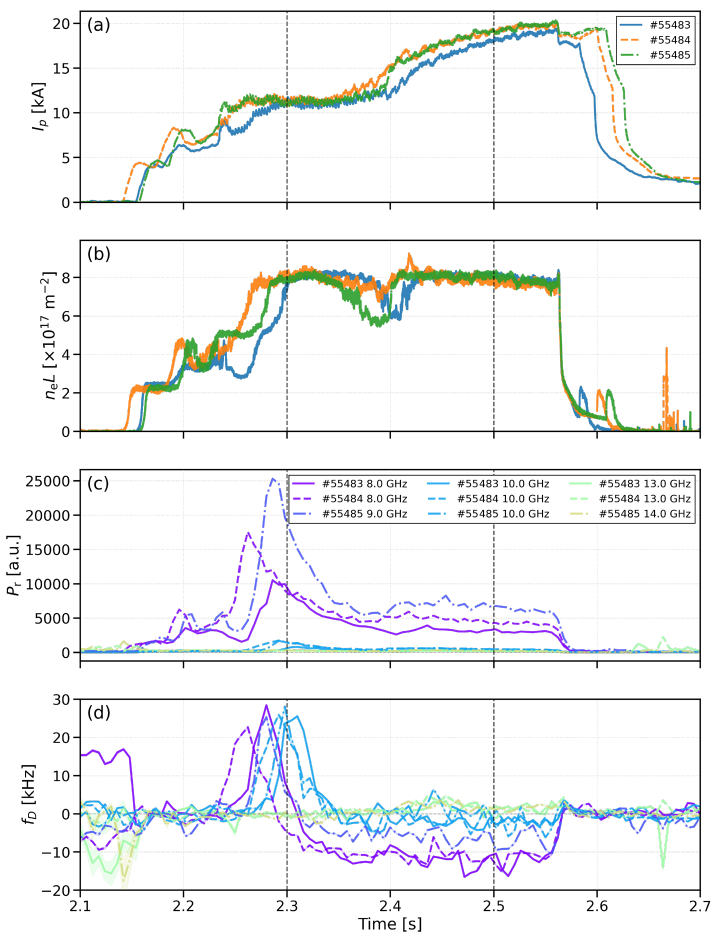}
\caption{Temporal evolution of the plasma parameters and DBS signals for the three reproducible discharges (\#55483, \#55484, and \#55485). 
(a) Plasma current $I_p$, (b) line-integrated electron density $n_e L$, (c) frequency-integrated DBS backscattered power intensity $P_r$, and (d) Doppler frequency shift $f_D$. 
In panels (a) and (b), the distinct colors correspond to different discharges. 
In panels (c) and (d), the colors represent different probing frequencies, while the line styles distinguish the different discharges.}
\label{fig:exp_res}
\end{figure}

Figures~\ref{fig:exp_res}(a) and (b) show the temporal evolution of the plasma current $I_{\rm p}$ and the line-integrated electron density $n_{\rm e}L$, respectively. 
The high reproducibility of these discharges is confirmed by the nearly identical temporal evolutions of both the plasma current, $I_{\rm p}$, and the line-integrated electron density, $n_{\rm e}L$, across the three selected shots. 
To illustrate the DBS responses under different plasma conditions, the DBS spectra at two representative time points, $t = 2.3$~s and $t = 2.5$~s, are presented. 
At $t = 2.3$~s, the plasma current was approximately $I_{\rm p}=10$~kA, while the line-integrated electron density was $n_{\rm e}L \approx 8 \times 10^{17}$~m$^{-2}$.
At $t = 2.5$~s, the plasma current increased to approximately $I_{\rm p}=20$~kA, whereas the line-integrated electron density remained nearly unchanged.
Representative DBS scattering spectra obtained at these two time points are shown in Fig.~\ref{fig:exp_res}, together with the corresponding Gaussian fitting results used to evaluate the DBS signal parameters.
\begin{figure}[t]
\centering
\includegraphics[width=\columnwidth]{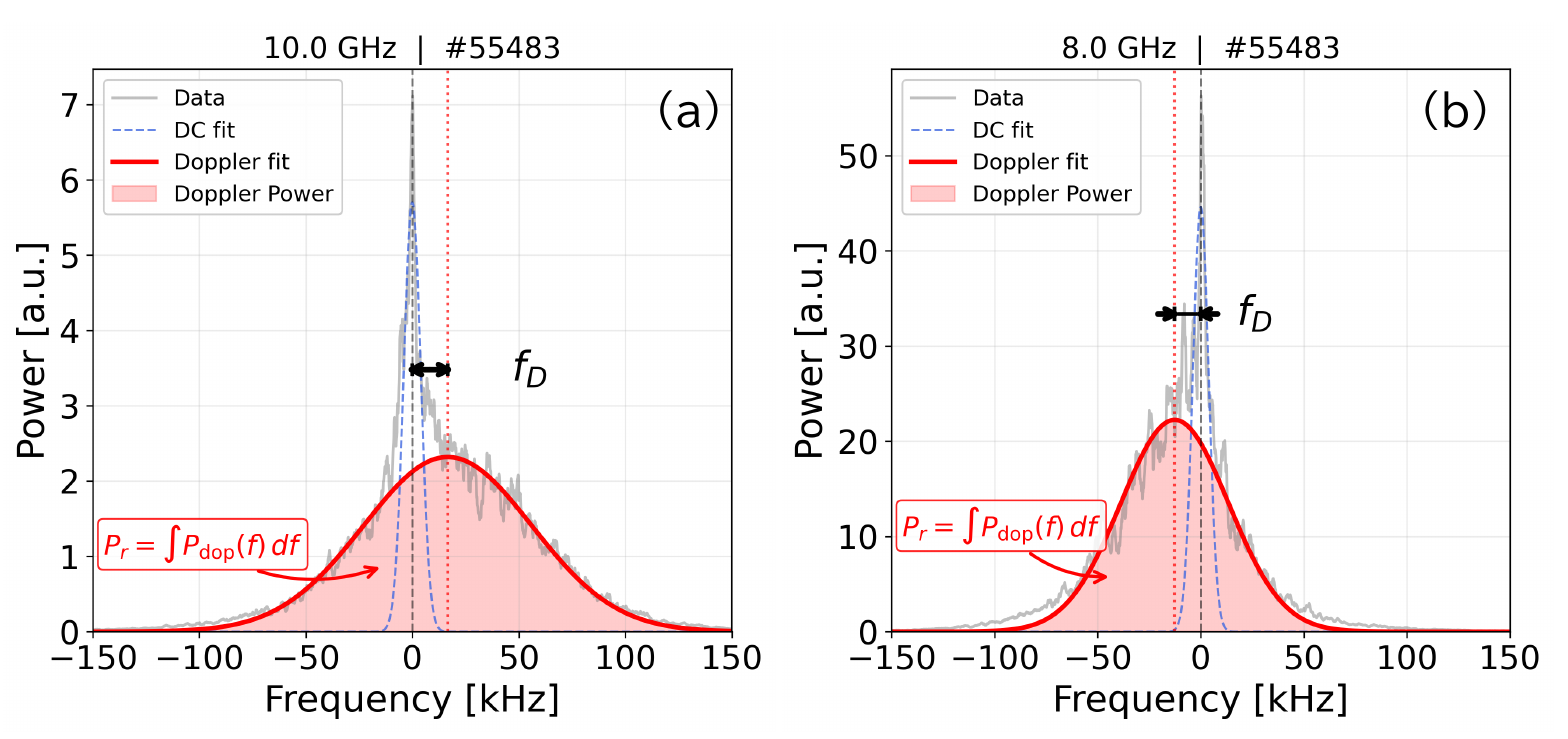}
\caption{
Representative DBS spectra measured at (a) $t=2.3$~s for the 10~GHz channel and (b) $t=2.5$~s for the 8~GHz channel.
The red curves indicate the Gaussian fits applied to the Doppler-shifted scattering component after removal of the zero-frequency contribution.
The scattered signal intensity $P_r$ was evaluated as the integrated area of the fitted Doppler-shifted component, while the Doppler frequency shift $f_D$ was determined from the center frequency of the Gaussian fit.
}\label{fig:exp_spec}
\end{figure}
The scattered signal intensity $P_r$ is evaluated as the integrated area under the Gaussian fit after the removal of the symmetric zero-frequency component, which primarily originates from stray light or unshifted background reflections. 
Additionally, the Doppler frequency shift $f_{\rm D}$ is directly determined from the center frequency of the fitted Gaussian profile.
By applying this fitting procedure continuously over the entire discharge duration, the full temporal profiles of the DBS responses are obtained, as displayed in Figs.~\ref{fig:exp_res}(c) and (d). 
The frequency-integrated scattered signal intensity reached its maximum around $t \sim 2.3$~s, accompanied by a distinct positive Doppler frequency shift. 
Among the multiple channels, the largest signal levels were predominantly observed at the lower probing frequencies of 8 and 9~GHz. 
Subsequently, as the discharge progressed, the signal intensity gradually decreased while the magnitude of the Doppler frequency shift approached 0~kHz. 
By $t = 2.5$~s, negative Doppler frequency shifts were clearly observed in the 8 and 9~GHz channels, whereas the higher-frequency channels remained close to 0~kHz.
These contrasting experimental observations indicate that the plasma and turbulence characteristics differ significantly between the two time points ($t = 2.3$~s and $t = 2.5$~s). 
In the present study, the focus is placed on obtaining a quantitative estimate of the density fluctuation amplitude using the proposed signal correction method rather than on discussing the underlying plasma physics. As an example, the profiles at $t = 2.5$~s are analyzed. 
The signal correction procedure formulated in the previous section is applied to remove the geometric and optical diagnostic response effects from the raw DBS data.

\subsection{\label{sec:calibration}Estimation of turbulence characteristic by signal correction}

\begin{figure}[t]
\centering
\includegraphics[width=\columnwidth]{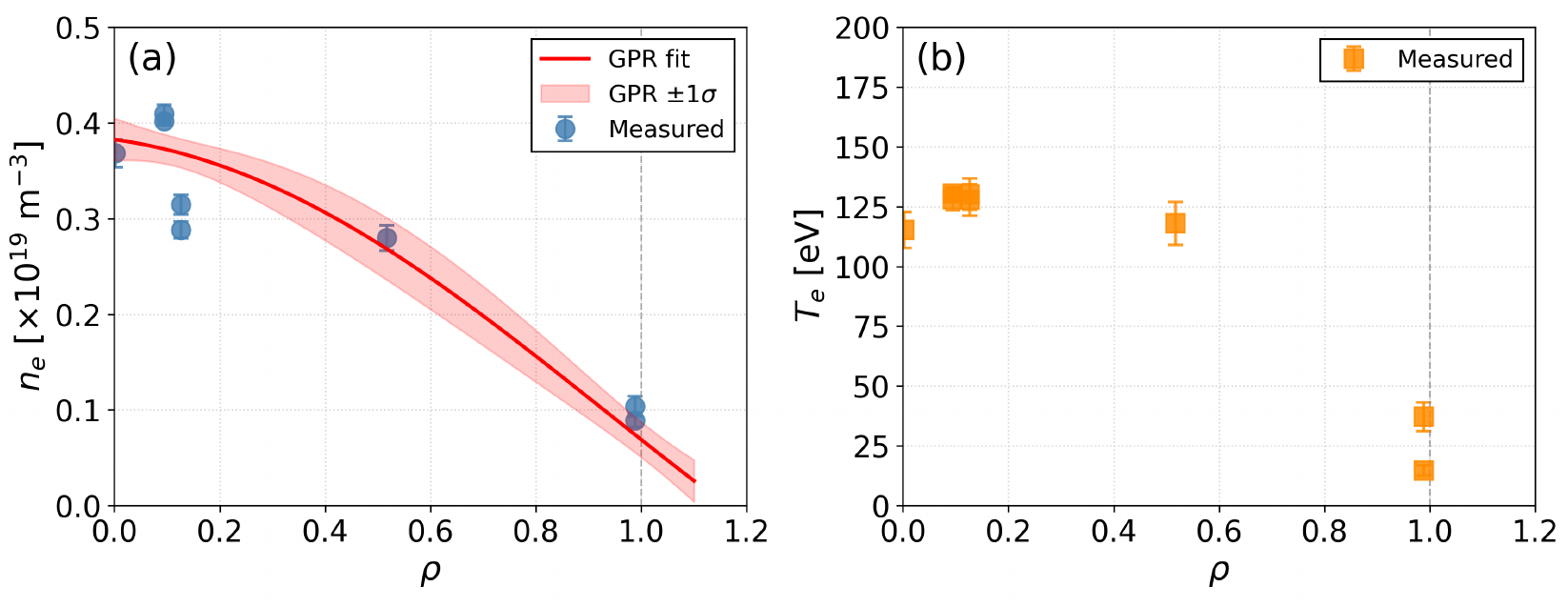}
\caption{
(a) Electron density and (b) electron temperature profiles at $t = 2.5$~s. The solid lines represent the fitted profiles, and the shaded regions denote the conservatively evaluated uncertainties. The symbols represent all Thomson scattering measurements obtained from shots \#55483--55485.
}
\label{fig:exp_neTe}
\end{figure}
\begin{figure}[t]
  \centering
  \includegraphics[width=\linewidth]{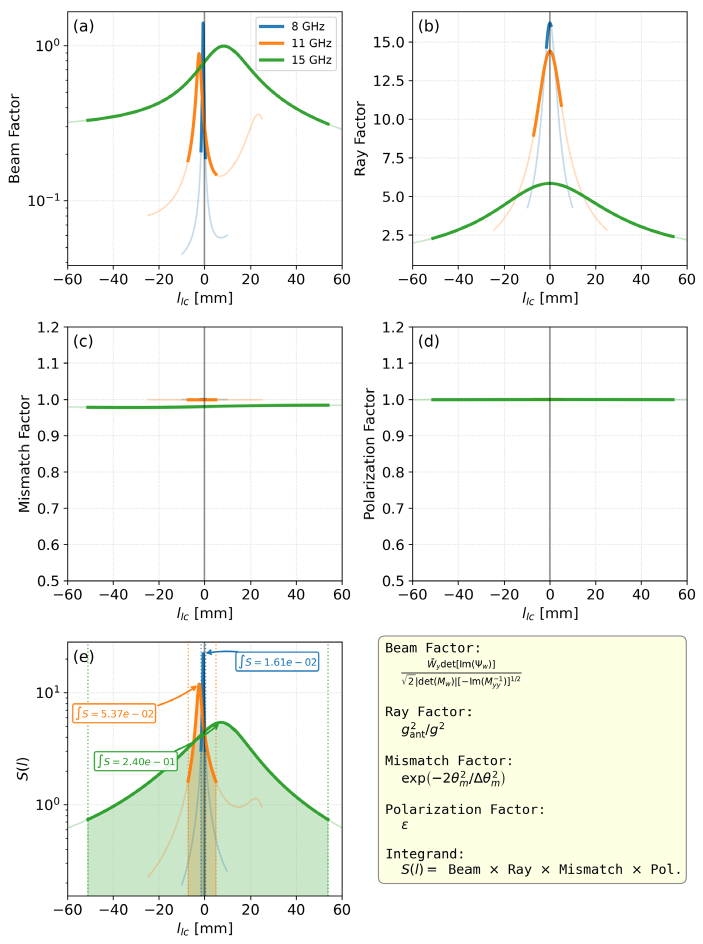}
  \caption{Calculated sensitivity function components as a function of the local ray coordinate $l_{\rm lc}$: (a) beam factor, (b) ray factor, (c) mismatch factor, (d) polarization factor, and (e) the resulting sensitivity function, $S(l)$. The bold segments indicate the effective scattering region bounded by $L_1$ and $L_2$, as determined from the sensitivity function shown in panel (e).
}
\label{fig:exp_sctparam}
\end{figure}
\begin{figure}[t]
\centering
\includegraphics[width=\linewidth]{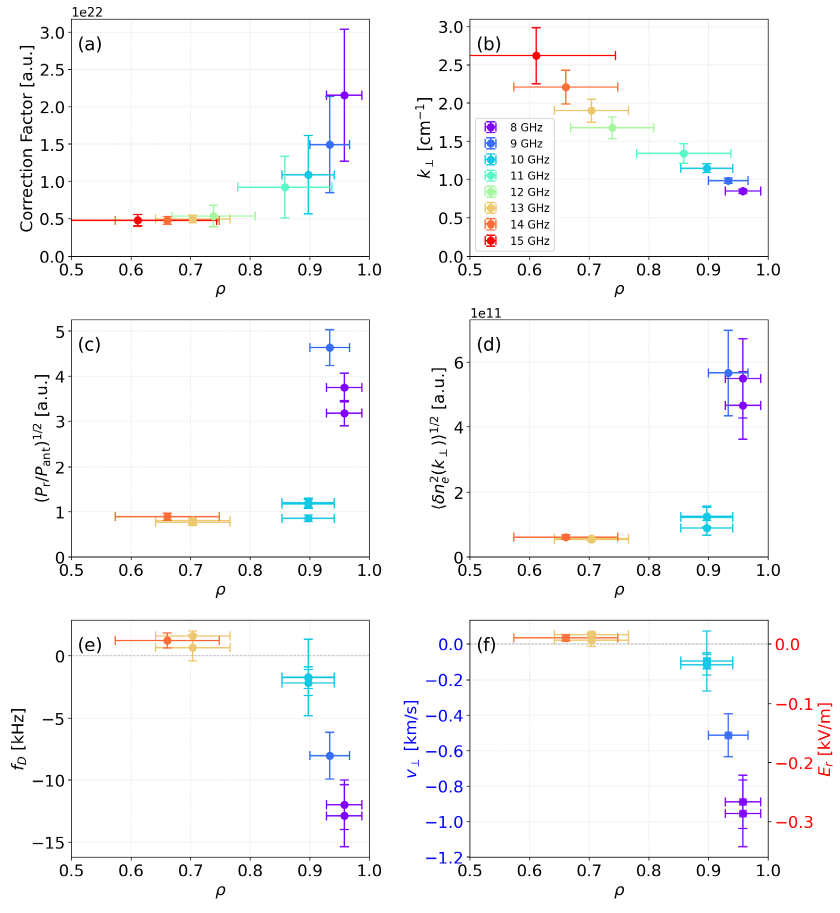}
\caption{Radial profiles of the quantities derived from the DBS analysis for the representative discharge at $t=2.5$~s: (a) correction factor $CF$, (b) perpendicular measurement wavenumber $k_\perp$, (c) uncorrected fluctuation amplitude $(P_r/P_{\rm ant})^{1/2}$, (d) corrected density fluctuation amplitude, (e) Doppler frequency shift $f_D$, and (f) estimated perpendicular velocity $v_\perp$ and radial electric field $E_r$. Horizontal error bars represent the measurement localization uncertainty, while vertical error bars denote the uncertainties propagated from the electron density profile fitting.}
\label{fig:exp_amp_Er}
\end{figure}

Figure~\ref{fig:exp_neTe} shows the electron density and electron temperature profiles at $t = 2.5$~s. Although the QUEST Thomson scattering diagnostic currently provides a total of eight spatial channels, only about four measurement points were located within the small plasma region analyzed in this study\cite{kono2023development}. 
Because these sparse measurements are insufficient to uniquely constrain the continuous electron density profile required for beam tracing, the $n_e$ profile is fitted using Gaussian process regression (GPR). 
The shaded region in Fig.~\ref{fig:exp_neTe} represents the one-standard-deviation (1$\sigma$) uncertainty estimated by GPR. 
To evaluate the influence of this uncertainty on the beam propagation analysis, 50 smooth electron density profiles were generated from the GPR posterior distribution and individually used as inputs to SCOTTY. 
The resulting ensemble of beam-tracing calculations was then used to estimate the uncertainty in the propagation characteristics. 
Figure~\ref{fig:exp_sctparam} shows a representative example of the calculated sensitivity function components along the local ray coordinate, $l_{\rm lc}$, for probing frequencies of 8, 11, and 15~GHz.
As shown in Figs.~\ref{fig:exp_sctparam}(a) and (b), both the beam and ray factors exhibit pronounced peaks near the cutoff location ($l_{\rm lc} = 0$) at lower frequencies, indicating a strong spatial localization of the scattering signal. 
With increasing probing frequency, however, these peaks broaden substantially and decrease in amplitude; at 15~GHz, the ray factor peak is reduced by more than a factor of two, resulting in a considerably flatter profile. 
This coordinated broadening and flattening demonstrate that the diagnostic sensitivity becomes progressively distributed over a wider region along the ray trajectory at higher frequencies, rather than remaining concentrated near the cutoff layer.
The mismatch factor shown in Fig.~\ref{fig:exp_sctparam}(c) exhibits only a weak dependence on the probing frequency, with variations remaining below 5\% over the frequency range considered. Similarly, the polarization factor [Fig.~\ref{fig:exp_sctparam}(d)] remains unity across all probing frequencies because the present measurements were performed exclusively in the O-mode.
In the present analysis, mode conversion between the O-mode and X-mode during propagation is neglected.
The resulting sensitivity function, $S(l)$, obtained as the product of the four components shown in Fig.~\ref{fig:exp_sctparam}(a)--(d), is shown in Fig.~\ref{fig:exp_sctparam}(e).
The bold segments indicate the effective scattering region bounded by $L_1$ and $L_2$.
As the probing frequency increases, the width of this region expands substantially due to the broadening of the beam and ray factors.
This behavior can be understood as follows. At higher probing frequencies, the ray penetrates deeper into the plasma and experiences weaker refraction in the cutoff region, resulting in a more gradual curvature over an extended propagation length. Consequently, even though the underlying density gradient remains unchanged, the contribution along the ray path does not remain localized but is distributed over a wider region. This leads to an apparent broadening of the effective scattering region.
On the other hand, when this broadening is mapped onto the $\rho$ coordinate, its effect is partially mitigated. While the spatial resolution degradation is not completely removed, it is less severe than that suggested in ray-space representation, appearing as an approximately several-fold increase (around a factor of five) compared with neighboring conditions (see Fig.~\ref{fig:exp_amp_Er}).
As a result, the integrated sensitivity increases from $1.61\times10^{-2}$ at 8~GHz to $2.40\times10^{-1}$ at 15~GHz.
This approximately fifteenfold increase demonstrates that frequency-dependent corrections are essential for the quantitative evaluation of density fluctuation amplitudes in QUEST.

Figure~\ref{fig:exp_amp_Er} summarizes the radial profiles of the correction factor evaluated using SCOTTY, the corresponding measurement wavenumber, the fluctuation amplitudes before and after correction, and the Doppler-derived plasma parameters.
Figure~\ref{fig:exp_amp_Er}(a) shows the radial profile of the correction factor. The correction factor exhibits a strong radial dependence and increases by approximately a factor of four toward the plasma edge ($\rho\sim0.95$) compared with the inner region ($\rho\sim0.6$). This behavior originates from the frequency dependence of the integrated sensitivity, $\int S(l) dl$, discussed in the previous section. As the probing frequency increases, the sensitivity profile becomes broader, resulting in a smaller correction factor. The horizontal error bars represent the uncertainty in the measurement location arising from the finite width of the sensitivity function together with the uncertainty in the fitted electron density profile. The vertical error bars were estimated by repeating the SCOTTY calculations using density profiles sampled from the GPR uncertainty. Because the launched microwave power $P_{\rm ant}$ was not monitored during the experiment, an additional uncertainty corresponding to $P_{\rm ant}=250$--350~mW, determined from subsequent bench-top measurements, was also included. No systematic frequency dependence of $P_{\rm ant}$ was observed in these measurements.
Figure~\ref{fig:exp_amp_Er}(b) shows the corresponding radial profile of the perpendicular wavenumber, $k_\perp$, which increases toward the plasma core from approximately $0.9$ to $2.6~{\rm cm^{-1}}$. Although the SCOTTY calculations were carried out at 1-GHz intervals, experimental data were available only at 8, 9, 10, 13, and 14~GHz.
Figure~\ref{fig:exp_amp_Er}(c) presents the normalized received signal amplitude, $(P_r/P_{\rm ant})^{1/2}$, which is proportional to the density fluctuation amplitude according to Eq.~(\ref{eq:fluctuation_correction}) and therefore represents the uncorrected fluctuation level. The largest amplitudes are observed in the edge region corresponding to the 8 and 9~GHz channels, whereas the signal decreases toward the plasma core.
After applying the correction factor, the corrected fluctuation amplitude shown in Fig.~\ref{fig:exp_amp_Er}(d) exhibits an even stronger edge localization. The enhancement near $\rho\sim0.95$ becomes more pronounced because the correction factor itself is largest in this region, indicating that the density fluctuations are strongly localized near the plasma edge in the present discharge.
Figure~\ref{fig:exp_amp_Er}(e) shows the radial profile of the Doppler frequency shift, $f_D$. Negative Doppler shifts are observed near the plasma edge, whereas the frequency shift remains close to zero in the inner region. The corresponding perpendicular velocity, $v_\perp$, and radial electric field, $E_r$, are presented in Fig.~\ref{fig:exp_amp_Er}(f). 
A finite negative radial electric field is formed near the plasma edge, while $E_r$ remains close to zero in the inner region. Notably, the region of enhanced fluctuation amplitude coincides with the formation of the negative radial electric field, suggesting a close relationship between the edge turbulence and the radial electric field structure. A detailed investigation of the underlying physics is beyond the scope of the present paper, whose primary objective is to demonstrate the proposed correction methodology.

\section{\label{sec:conclusion}Conclusion}
A three-channel X-/Ku-band DBS system has been developed and installed on QUEST for turbulence and electric-field measurements. 
To enable the quantitative evaluation of density fluctuation amplitudes, a correction methodology based on the synthetic DBS code SCOTTY was established. 
The diagnostic response, including beam propagation, ray amplification, magnetic mismatch, and polarization effects, was evaluated along the beam trajectory and incorporated into a comprehensive correction factor. 
The analysis showed that the diagnostic sensitivity exhibits a strong frequency dependence, leading to significant variations in both the integrated sensitivity and the correction factor over the measurement range. 
These results demonstrate that correcting for the diagnostic response is essential for quantitatively comparing fluctuation amplitudes measured at different probing frequencies.
By applying the proposed correction factor to the measured scattering signals, quantitative density fluctuation amplitudes were successfully reconstructed. 
The corrected profiles revealed enhanced fluctuation activity near the plasma edge, where a finite negative radial electric field was also observed. 
The proposed methodology provides a practical framework for quantitative DBS measurements in spherical tokamaks, where microwave propagation effects play a particularly important role.

This work represents the first quantitative evaluation of density fluctuations using low-frequency X-/Ku-band DBS measurements in QUEST. 
The proposed SCOTTY-based correction methodology provides a practical framework for quantitative DBS measurements in spherical tokamaks and establishes a foundation for future studies of turbulence, transport, and aspect-ratio-dependent confinement physics.

\begin{acknowledgments}
This work was supported by the Foundation of Kinoshita Memorial Enterprise, JSPS KAKENHI Grant Number 24K00613, the NIFS Collaboration Research Program (24KUTR190), and 'A*STAR STRIDE grant (H26-MSE152).
\end{acknowledgments}

\bibliography{refbib}

@article{hanada2025experimental,
  title = {Experimental progress and future plans on spherical tokamak {QUEST}},
  author = {Hanada, K and Idei, H and Ido, T and Ikezoe, R and Nagashima, Y and Hasegawa, M and Onchi, T and Kinoshita, T and Kuroda, K and Oya, M and others},
  journal = {Plasma Physics and Controlled Fusion},
  volume = {67},
  number = {11},
  pages = {115031},
  year = {2025},
  publisher = {IOP Publishing}
}

@article{hall2022beam,
  title = {Beam model of Doppler backscattering},
  author = {Hall-Chen, Valerian H and Parra, Felix I and Hillesheim, Jon C},
  journal = {Plasma Physics and Controlled Fusion},
  volume = {64},
  number = {9},
  pages = {095002},
  year = {2022},
  publisher = {IOP Publishing}
}

@article{kaye2021thermal,
  title = {Thermal confinement and transport in spherical tokamaks: a review},
  author = {Kaye, S. M. and Connor, J. W. and Roach, C. M.},
  journal = {Plasma Physics and Controlled Fusion},
  volume = {63},
  number = {12},
  pages = {123001},
  year = {2021},
  publisher = {IOP Publishing}
}

@article{doyle2007chapter,
  title = {Chapter 2: Plasma confinement and transport},
  author = {Doyle, E. J. and Houlberg, W. A. and Kamada, Y. and Mukhovatov, V. and Osborne, T. H. and Polevoi, A. and Bateman, G. and Connor, J. W. and Cordey, J. G. and Fujita, T.},
  journal = {Nuclear Fusion},
  volume = {47},
  number = {6},
  pages = {S18--S127},
  year = {2007}
}

@article{kono2023development,
  title = {Development of Thomson scattering measurement system for long duration discharges on the {QUEST} spherical tokamak},
  author = {Kono, Kaori and Ido, Takeshi and Ejiri, Akira and Hanada, Kazuaki and Yue, Qilin and Hasegawa, Makoto and Peng, Yi and Sakai, Seiya and Ikezoe, Ryuya and Idei, Hiroshi and others},
  journal = {Plasma and Fusion Research},
  volume = {18},
  pages = {1405012},
  year = {2023},
  publisher = {The Japan Society of Plasma Science and Nuclear Fusion Research}
}

@article{kohagura2022ku,
  title = {Ku-band multichannel frequency comb Doppler reflectometer on the {GAMMA 10/PDX} tandem mirror},
  author = {Kohagura, J. and Tokuzawa, T. and Yoshikawa, M. and Shima, Y. and Nakanishi, H. and Nakashima, Y. and Sakamoto, M. and Katoh, H.},
  journal = {Review of Scientific Instruments},
  volume = {93},
  number = {12},
  year = {2022},
  publisher = {AIP Publishing}
}

@article{pratt2024density,
  title = {Density wavenumber spectrum measurements, synthetic diagnostic development, and tests of quasilinear turbulence modeling in the core of electron-heated {DIII-D} H-mode plasmas},
  author = {Pratt, Q. and Hall-Chen, V. and Neiser, T. F. and Hong, R. and Damba, J. and Rhodes, T. L. and Thome, K. E. and Yang, J. and Haskey, S. R. and Cote, T. and others},
  journal = {Nuclear Fusion},
  volume = {64},
  number = {1},
  pages = {016001},
  year = {2024},
  publisher = {IOP Publishing}
}

\end{document}